\documentclass[9pt,onecolumn,twoside]{osajnl}

\journal{ol} 

\setboolean{shortarticle}{false}

\usepackage[T1]{fontenc}
\usepackage[utf8]{inputenc}
\usepackage{amsmath,amssymb}
\usepackage{graphicx}
\usepackage{multirow} 
\usepackage{subfigure} 

\usepackage{lineno}

\begin{document}


\title{Integrated Photonic Programmable Random Matrix Generator with Minimal Active Components}

\author[1]{Kevin Zelaya}
\author[1,2]{Mostafa Honari-Latifpour}
\author[1,2,*]{Mohammad-Ali Miri}

\affil[1]{Department of Physics, Queens College of the City University of New York, Queens, New York 11367, USA}
\affil[2]{Physics Program, The Graduate Center, City University of New York, New York, New York 10016, USA}

\affil[*]{Corresponding author: mmirilab@gmail.com}

\begin{abstract}
Random matrices are fundamental in photonic computing because of their ability to model and enhance complex light interactions and signal processing capabilities. In manipulating classical light, random operations are utilized for random projections and dimensionality reduction, which are important for analog signal processing, computing, and imaging. In quantum information processing, random unitary operations are essential to boson sampling algorithms for multiphoton states in linear photonic circuits. In photonic circuits, random operations are realized through disordered structures resulting in fixed unitary operations or through large meshes of interferometers and reconfigurable phase shifters, which require a large number of phase shifters. In this article, we introduce a compact photonic circuit for generating random matrices by utilizing programmable phase modulation layers interlaced with a fixed mixing operator. We show that using only two random phase layers is sufficient for producing output optical signals with a white-noise profile, even for highly sparse input optical signals. We experimentally demonstrate these results using a silicon photonics circuit with tunable thermal phase shifters and utilize waveguide lattices as mixing layers. The proposed circuit offers a practical method for generating random matrices for photonic information processing and for applications in data encryption.
\end{abstract}

\maketitle

\section{Introduction}




The rapid advancements in photonics fabrication techniques and materials science have enabled the deployment of photonic integrated circuits compatible with telecommunication wavelengths in compact on-chip form factors~\cite{carolan2015universal, bogaerts2020programmable, Bogaerts20b}. This allows exploiting the properties of light to perform computational tasks with reduced power consumption, increased bandwidth, and improved reliability. The ability to manipulate light on chip-scale platforms has sparked a plethora of applications across various fields, including telecommunications~\cite{paraiso2021photonic}, optical neural networks~\cite{shen2017deep,lin2018all} and machine learning~\cite{li2024high}, sensing~\cite{chrostowski2012silicon}, imaging~\cite{wang2024integrated}, and quantum computing~\cite{harris2017quantum, Wang2019}. Indeed, extensive research has been developed to deploy programmable photonic integrated circuits capable of real-time tuning. The idea of such devices was first introduced by Reck \textit{et al.}~\cite{reck1994experimental} for free-space propagation using arrays of beam splitters and phase shifters, which paved the way for compact on-chip solutions based on meshes of Mach-Zehnder interferometers (MZI) \cite{miller2013self, miller2015perfect, clements2016optimal, hamerly2022accurate, Bogaerts20a, dai2024programmable}, as well as recirculating meshes \cite{zhuang2015programmable, perez2017multipurpose, perez2019integrated}, and multi-plane light conversion and multiport waveguide arrays~\cite{taguchi2023iterative,tanomura2020robust,markowitz2023universal, Markowitz23Auto, zelaya2024goldilocks, pastor2021arbitrary}.

Recently, interlaced architectures that represent arbitrary unitary $N\times N$ matrices have been explored as alternative candidates to conventional MZI meshes~\cite{markowitz2023universal, Markowitz23Auto, zelaya2024goldilocks}, where arrays of phase shifters and passive mixing layers are intertwined one after the another. This approach involves the intertwining of $N+2$ passive and $N+1$ active layers of optical elements, which has demonstrated high flexibility. Specifically, the passive layer responsible for mixing light propagation across all waveguides does not need to follow any specific design as long as its corresponding transfer matrix satisfies a density criterion~\cite{zelaya2024goldilocks}. The active elements lie in a layer separated from the passive one, allowing for more flexibility in the phase elements, which can be implemented through microheaters~\cite{harris2014efficient,liu2022thermo}, phase-change materials~\cite{rios2022ultra}, or other technologies.

Along different lines, the implementation of random unitary matrices by purely optical means has found exciting applications \cite{gigan2022imaging}, such as optical encryption in free-space settings~\cite{refregier1995optical,frauel2007resistance} and through metasurfaces~\cite{audhkhasi2024full,audhkhasi2024leveraging}. Harnessing the random transmission matrix of light in disordered waveguide arrays allows for encoding high-dimensional images into their lower-dimensional representations ~\cite{miri2021integrated,wang2024integrated}. This property has been found resourceful in random disordered fibers, where the inherent Anderson localization renders highly localized modes used for high-fidelity transport of intensity patterns and images~\cite{mafi2019disordered, mafi2021review, cao2022harnessing, segev2013anderson, abouraddy2012anderson, dikopoltsev2022observation}. Furthermore, random photonic devices have been shown to be an excellent resource for generating operations akin to Haar-random matrices, a fundamental task on boson sampling required in quantum computing tasks~\cite{russell2017direct, clementi2023programmable}.

The present work introduces a programmable, compact, and simple-to-fabricate photonic chip designed to randomize light signals injected into its input. Such a process is achieved by exploiting general interlaced architectures discussed in the literature~\cite{Markowitz23Auto,zelaya2024goldilocks} and reducing the corresponding number of elements to an effective minimum without compromising its functionality. Indeed, numerical experiments reveal that only two active layers are required to perform the randomization process with high accuracy. Here, the output randomness is inherited from the random distributions assigned to the phase elements; the quality of the random output is evaluated by comparing it with the typical profile of white noise signals. Although the fabricated chip randomizes both the real and imaginary parts of the input signal, we present a relatively simple scheme in which the random patterns are still measurable with conventional power measurements. The chip capabilities are extended by demultiplexing large signals into smaller sizes without jeopardizing the quality of the random output. Furthermore, a specific application is introduced, where the chip can be used as an all-optical encryption device whose decryption process can be assessed through an equivalent device, the existence of which is guaranteed by the unitary nature of the randomization device.


\section{Results}
\begin{figure*}[t]
    \centering
    \includegraphics[width=0.75\textwidth]{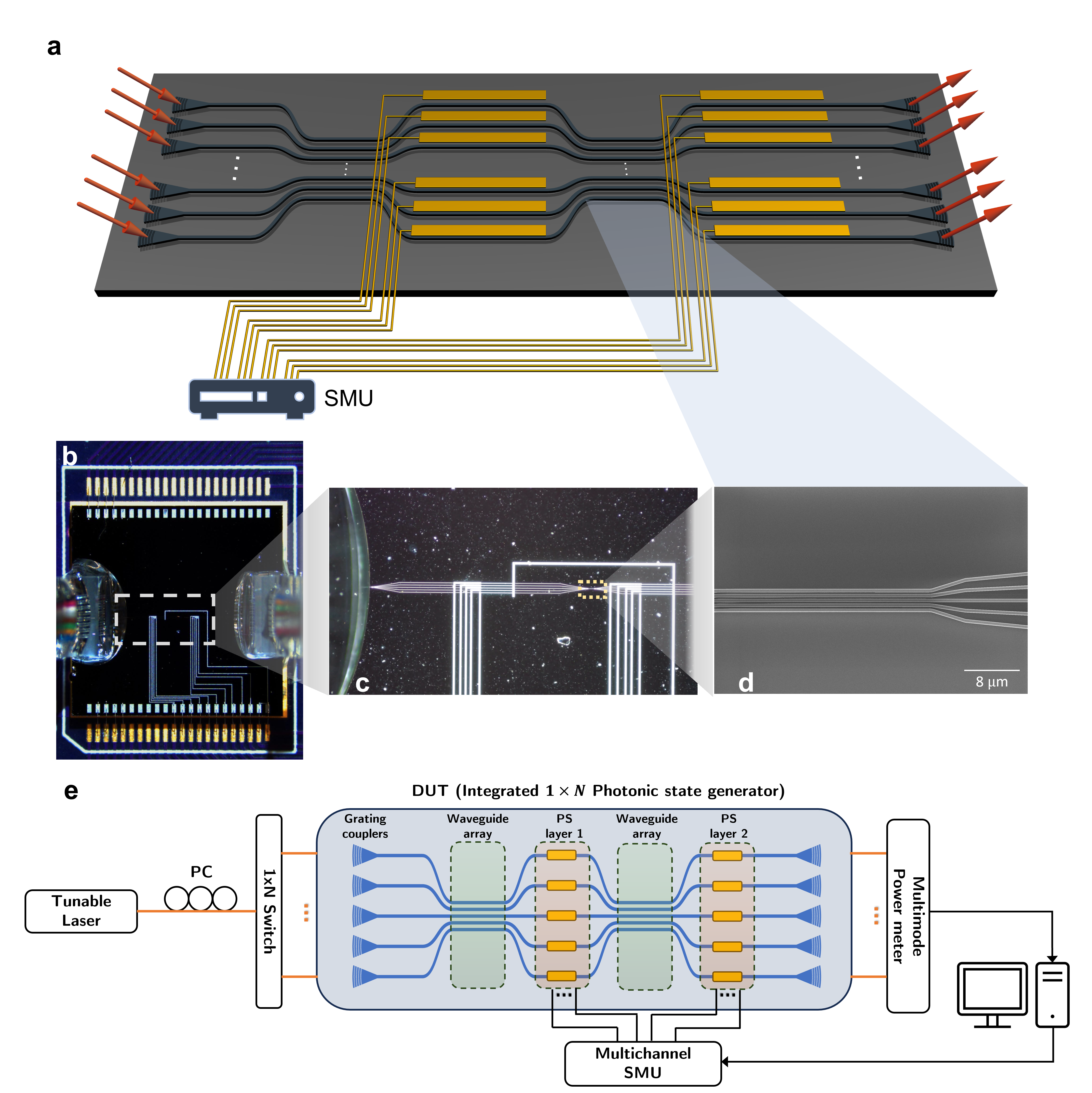}
    \caption{\textbf{Experimental setup and PIC design.} \textbf{a} PIC interlaced structure for $M=2$ layers. The input optical signal $\mathbf{x}$ is fed into the PIC, and the randomization phases are programmed by the SMU controller. The processed optical output signal is $\mathbf{z}$. For completeness, the fully packaged fabricated chip (\textbf{b}), a microscope image of the photonic circuit area (\textbf{c}), and a SEM capture of the waveguide array section (\textbf{d}) are illustrated. \textbf{e} Experimental setup and operation of the randomization device. Here, the programmable photonic integrated circuit performing the randomization is set as the device under test (DUT). In contrast, the phase shifters are the programmable elements of the device, which are externally and individually controlled by a multichannel source measure unit (SMU). The processed light at the output grating couplers is collected through a multiport power meter.
    }
    \label{fig:block}
\end{figure*}

\subsection{Random architecture model}
Photonic integrated circuits (PICs) are rapidly becoming an attractive solution for optical computing applications. Their versatility enables the design of both programmable units~\cite{zelaya2024goldilocks} and integrated task-specific operations~\cite{meng2023compact,zelaya2024integrated}. Particularly, unitary architectures based on interlaced layers of passive and active optical components have become a common design solution, for they allow for the representation of arbitrary unitary matrices (universality) and have shown to be resilient to manufacturing errors~\cite{Markowitz23Auto,taguchi2024standalone}. The fundamental operational principle lies in the propagation of guided modes through waveguides $\mathbf{E}(\mathbf{r})=\mathcal{E}(\mathbf{r}_{\perp})e^{i(wt-\beta \mathbf{r}_{\parallel})}\hat{e}_{\perp}$, with $\mathcal{E}(\mathbf{r}_{\perp})$ the normalized guided-mode amplitude, $r_{\parallel}$ and $\mathbf{r}_{\perp}$ position vectors in the direction parallel and perpendicular to the propagation, respectively, $\hat{e}_{\perp}$ the unit vector in the perpendicular direction, and $\beta$ the corresponding mode propagation constant~\cite{yariv2007photonics} (see Supplementary Material S1 for more details). The electric field of a $N$-port device composed of single-mode waveguides, such as the one illustrated in Figure~\ref{fig:block}\textbf{a}, writes as the complex-valued vector $\mathbf{x}=(x_{1},\ldots,x_{N})\mathcal{E}(\mathbf{r}_{\perp})$, with $x_{i}\in\mathbb{C}^{N}$ carrying information about the intensity and phase of the propagating light. Henceforth, to reduce the notation, the electric field propagating through the device is simply written as $\mathbf{x}\equiv(x_{1}\ldots,x_{N})$.

In general, universal interlaced architectures are composed of $M_{1}$ layers of passive mixing components and $M_{2}$ layers of active elements~\cite{Markowitz23Auto,zelaya2024goldilocks}. Particularly, a universal $N$-port unitary device requires $M_{1}=N+2$ passive and $M_{2}=N+1$ active layers to render any arbitrary unitary optical operation. In turn, for randomization tasks, we require that the output vectors show traces of randomness regardless of the nature of the injected input signal, and thus, not all layers might be needed. That is, we seek the minimum number of layers $M_{1}$ and $M_{2}$ so that the output resembles a white noise signal. Indeed, as pointed out in~\cite{zelaya2024goldilocks}, the intermediate passive layers do not necessarily require to take any specific form, and layers described by transmission matrices with dense properties render the desired functionality.

In the present design, the $M$-layer ($M_{1}=M_{2}=M$) unitary interlaced structure $\mathcal{U}\in U(N)$ is built based on dense passive layers $F(\alpha_{m})=e^{i\alpha_{m}H}$, which rule the wave evolution of light and are described by coupled-mode theory approach~\cite{Huang94,cooper2009numerically}. Here, $H$ is a tri-diagonal matrix with components $H_{n,n+1}=\kappa_{n}$ for $n\in\{1,\ldots,N-1\}$ and $\kappa_{n}$ the corresponding coupling parameter between neighbor waveguides. The device is operated by preparing an arbitrary input state $\mathbf{x}\in\mathbb{C}^{N}$, the input signal, which is subsequently randomized through the unitary transformation
\begin{equation}
\mathbf{x}_{out}=\mathcal{U}\mathbf{x} , \quad \mathcal{U} = P^{(M)} F(\alpha_{M})  \ldots P^{(1)} F(\alpha_{1}) , \quad M\in\mathbb{N} , 
\label{encrypt}
\end{equation}
where $P^{(j)}=diag(e^{i \phi^{(j)}_1}, \ldots, e^{i \phi^{(j)}_N})$ are unitary diagonal matrices characterizing the programmable phase-mask layers, whereas $\alpha_{j}$ are the coupling lengths for each waveguide array, for $j\in\{1,\ldots, M\}$. 

The randomization process can be achieved with high accuracy by incorporating only two active layers of phase shifters, $M=2$, and no substantial improvement is observed when more layers are included (see discussion below). This effectively reduces the overall size of the final architecture, rendering a low-footprint solution that requires a minimum number of control elements and, thus, is less prone to operational errors. The coupling coefficients defining the waveguide array can be chosen as those of the DFrFT operation using the Jx lattice~\cite{Wei16,Markowitz23Auto}, homogeneous lattice~\cite{Chr03}, or any random lattice that fulfills the density criterion posed by the Goldilocks principle discussed in~\cite{zelaya2024goldilocks} (see Supplementary Materials~S2 for examples of dense matrices).


\subsection{Experimental setup and measurements}
\begin{figure*}[t]
    \centering
    \includegraphics[width=0.85\textwidth]{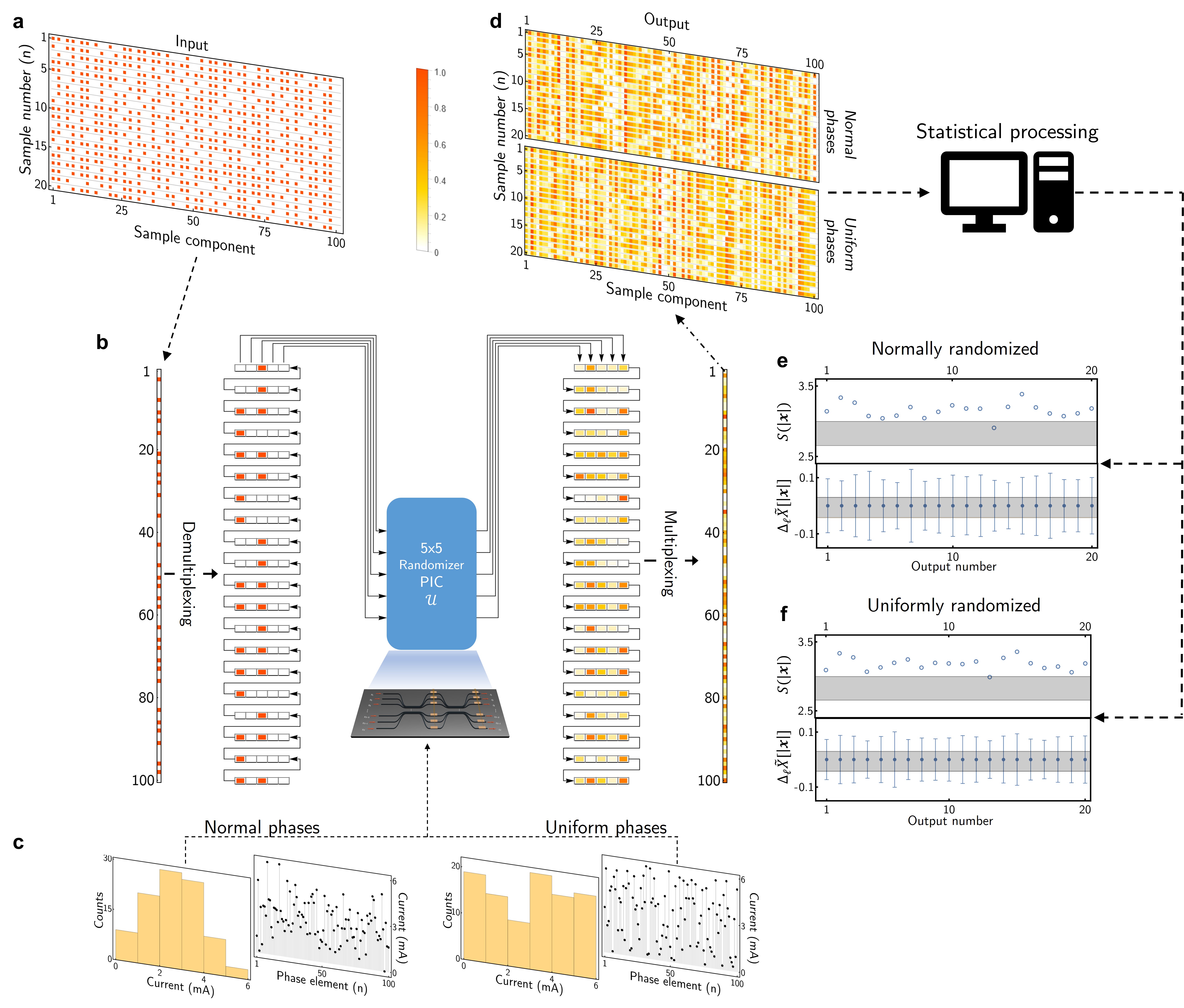}
    \caption{\textbf{Experimental run and data processing.} \textbf{a} Sequences of testing random pulsed trains $\mathbf{x}^{(p)}\in\mathbb{R}^{100}$, for $p\in\{1,\ldots,20\}$. \textbf{b} The latter are demultiplexed into signals $\hat{\mathbf{x}}^{(p)}:=\mathbb{R}^{20\times 5}$, which are programmed in the $N\times 1$ switch and injected into the PIC. This produces the randomized demultiplexed signals $\hat{\mathbf{z}}^{(p)}$. \textbf{c} In this process, two ensembles of random phases are loaded into the phase shifters through the SMU, which correspondingly powers the metal heaters. Such ensembles are shown as histograms and scatter plots, which highlight the normal (left) and uniform (right) distribution profiles. \textbf{d} The output processed signals from the PIC are multiplexed back to the vectors $\mathbf{z}^{(p)}\in\mathbb{R}^{100}$. \textbf{f}-\textbf{g} These outputs are then post-processed to extract the statistical information for normally and uniformly randomized phase distributions using the entropy $S[\mathbf{\vert x\vert}]$ and autocorrelation difference $\Delta_{\ell}\mathbf{X}[\vert \mathbf{x} \vert]$ criteria.}
    \label{fig:demux}
\end{figure*}
The PIC performing the randomization operation is fabricated on a silicon-on-silica (SOI) platform (see Methods for details). The design relies on coupled waveguides to perform the mixing layer operation and a layer of metal heaters producing the desired phase shift using conventional thermo-optic effects~\cite{liu2022thermo}. TE grating couplers are used to externally couple the PIC to the light input source and at the output for the data collection stage. Furthermore, a mechanical polarization controller (PC) is attached to the injection fiber, which is tuned so that maximum power is coupled to the PIC. The device under test (DUT) is the randomization PIC depicted in Figure~\ref{fig:block}\textbf{a}, fed by a $N\times 1$ network switch that splits the quasi-TE0 mode into the desired inputs that encode the optical signal. The heaters producing the phase shift are electronically controlled by a multichannel source measure unit, providing independent control currents of up to $10$ mA to each metal heater. Lastly, the output modes are gathered from the output grating couplers and collected at a multiport power meter.

The proposed PIC design is flexible enough to be manufactured in commercial foundries. For the current experimental run, the PIC was designed as a $5\times 5$ unitary device (5-port device) in the form of a fully electrically and optically packaged chip (see Figure~\ref{fig:block}\textbf{b-d}). The packaging allows for easy and minimum-error light coupling and light-detection processes. Figure~\ref{fig:block}\textbf{e} summarizes the current experimental setup.

Manufacturing highly dense photonic chips presents challenges due to thermal and optical crosstalk, which become significant design problems when scaling up photonic units. To address this, we reduce the required number of ports of the proposed PIC by demultiplexing the input signal into lower-dimensional vectors, which are then sequentially fed at the input ports. Each input sequence is randomized using phases that are randomly distributed from either a normal or uniform distribution. This results in independent random outputs. Finally, the sequentially randomized outputs are multiplexed back into a higher-dimensional signal. This design choice is two-fold: it helps mitigate errors and reduces the device footprint. 

Thus, any input signal $\mathbf{x}\in\mathbb{R}^{100}$ shall be demuxed into sequences of 5-dimensional vectors, $\mathbf{x}\in\mathbb{R}^{20\times 5}$. Particularly, we consider highly sparse 100-dimensional vectors, $\mathbf{x}\in \mathbb{R}^{100}$, whose components are either zero or one so that at least a component with a value of one exists for every demuxed sequence. The zero and one values denote in our experiment the cases when no light is injected and when light is coupled to the corresponding grating coupler, respectively. The distribution of zeros and ones is randomly selected for each sequence in the demuxed signal. The choice of such sparse signals is twofold: they can be implemented straightforwardly with the current setup, and they have poor qualities when it comes to random signals. The latter implies that if random footprints are found at the output of the PIC, it is not due to the random nature of the input signal (see analysis below). Since the PIC output is gathered through a power meter, any phase information is washed out during power measurements. Thus, real and imaginary parts of the complex-valued randomized output are not accessible through power detection schemes. Still, traces of randomness can be detected since the power measurements of complex-valued white noise (normally distributed signals) render a Rayleigh distribution instead~\cite{forbes2011statistical}.

The set of 20 input sparse signals used in the present experiment, the demuxed lower dimensional signals, and the corresponding measurements at the PIC output are illustrated in Figure~\ref{fig:demux}\textbf{a-c}. Furthermore, two runs are performed. In each run, the currents in the SMU are programmed in such a way that the corresponding phases follow normal and uniform distributions (Figure~\ref{fig:demux}\textbf{c}). That is, the set of sparse inputs is processed, and the corresponding output power is measured by loading one of the chosen phase distributions (Figure~\ref{fig:demux}\textbf{d}). The statistical properties of the gathered power measurements can be analyzed in order to determine the quality of randomness. Autocorrelation and Shannon entropy are two complementary criteria that allow qualifying the randomness of the output signal (see Section~2D for a thorough criteria analysis). Indeed, the output power measurements obtained by using either of the phase distributions are shown in  Figure~\ref{fig:demux}\textbf{e-f}). In the latter, the gray-shaded areas highlight the regions where white-noise behavior is expected in terms of power measurements (Rayleigh distribution), and the experimental results are shown for both normally and uniformly distributed phases. In both cases, the measured optical signals follow the expected random patterns with mild deviations in the entropy criteria. 

\subsection{White noise estimation}
\begin{figure*}[t]
    \centering
    \includegraphics[width=0.8\textwidth]{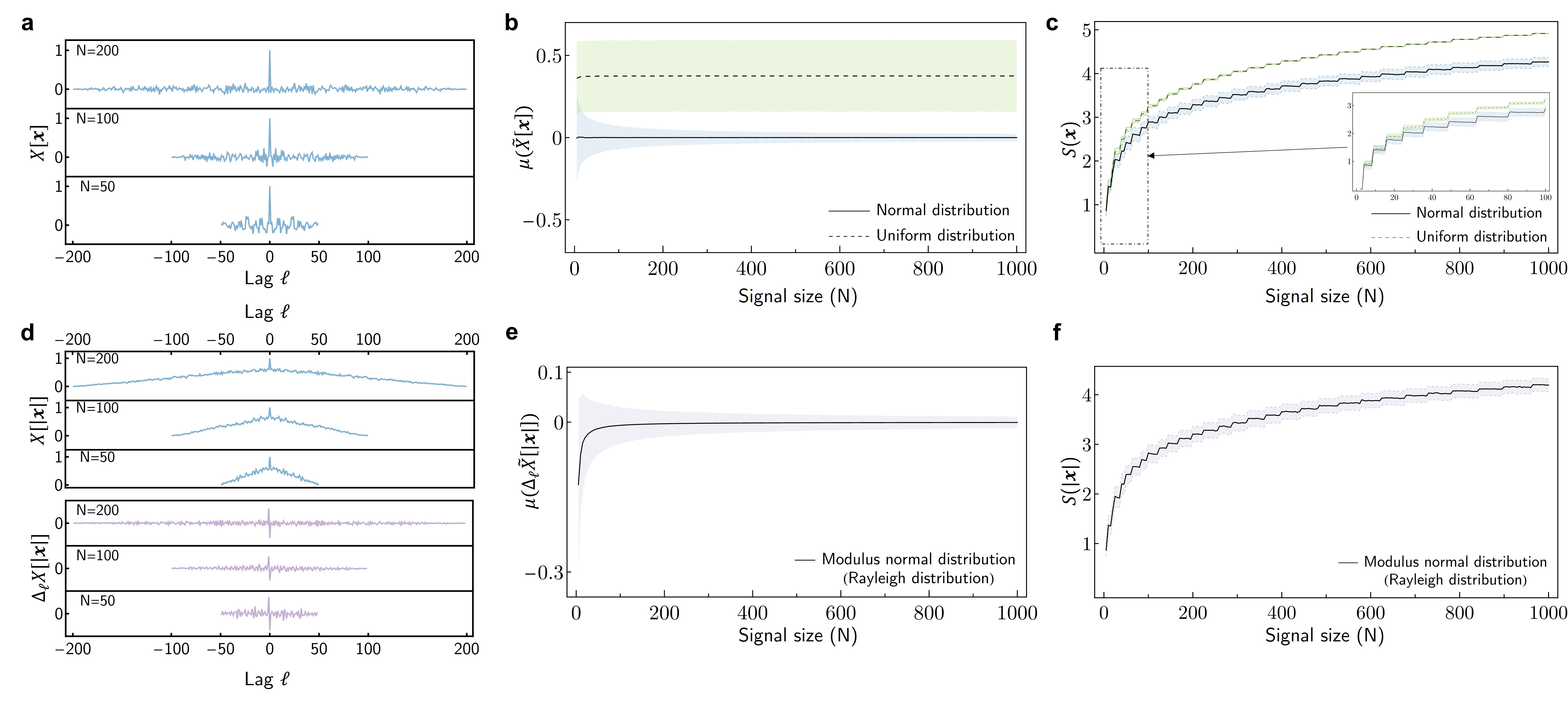}
    \caption{\textbf{White-noise criteria.} (a) Typical autocorrelation profile for white noise signals. (b) Truncated autocorrelation criterion $\widetilde{X}[\cdot]$ and (c) Shannon entropy ($S$) for ensembles of normal (blue-shaded) and uniform (purple-shaded) distributions as a function of the distribution size $N$. (d) Typical autocorrelation profile (upper panel) and the corresponding finite difference $\Delta_{\ell}$ of the modulus of white noise signals. The corresponding truncated autocorrelation criterion (e) and Shannon entropy (f) as a function of the distribution size $N$. In (b)-(c) and (e)-(f), 1000 random distributions were generated for each $N$, from which the mean (solid or dashed line) and standard deviations (shaded area) are computed.}
    \label{fig:entropy1}
\end{figure*}

To assess the randomness of the device outputs, some criteria must be established to classify any given signal as random white noise. Although this may be accomplished through several statistical measures, here we focus on the correlation and entropy properties. The use of two different statistical properties allows for ruling out false positives inherent in either autocorrelation or entropy analysis, as discussed below. White noise signals $\mathbf{x}$ are known to be uncorrelated to shifted copies of themselves, henceforth called lags and denoted by $\ell$. For continuous or infinitely sampled signals, the autocorrelation of white noise signals ($\mathbf{X}[\mathbf{x}]$) becomes a single impulse at the lag $\ell=0$, $X_{\ell}[\mathbf{x}]=\delta_{\ell,0}$ with $\delta_{p,q}$ the Kronecker-delta distribution. In turn, for finite discrete signals, the autocorrelation is approximately flat for $\ell\neq 0$ and peaks to the unity for $\ell=0$. Indeed, Figure~\ref{fig:entropy1}\textbf{a} depicts some typical profiles of $X[\bf{x}]$, for signals $\mathbf{x}\in\mathbb{R}^{N}$ normally distributed.

For $\bf{x}\in\mathbb{R}^{N}$, the autocorrelation renders a vector $\mathbf{X}[\mathbf{x}]\in\mathbb{R}^{2N-1}$, which is symmetric around the lag $\ell=0$; i.e., $X_{-\ell}[\mathbf{x}]=X_{\ell}[\mathbf{x}]$. For $\ell=0$, the autocorrelation reduces to the Euclidean norm of $\mathbf{x}$, $X_{\ell=0}[\mathbf{x}]=(\mathbf{x},\mathbf{x})=\Vert \mathbf{x} \Vert^{2}$. For normalized signals, the autocorrelation peaks at $\ell=0$ to the unity. Thus, throughout the manuscript, all signals analyzed by the autocorrelation are normalized beforehand. Following the autocorrelation symmetric, we focus exclusively on the positive lags $\ell>0$ and thus define the truncated autocorrelation
\begin{equation}
\widetilde{\mathbf{X}}[\mathbf{x}]:=\left( X_{1}, \ldots, X_{N-1} \right) ,
\label{chopped-xcorr}
\end{equation}
which contains the minimum relevant statistical information to be processed.

Further insight into the autocorrelation and entropy estimation can be achieved by considering two specific sets of signals, i.e., signals generated from the normal and uniform distribution. The normal distribution is the desired behavior for the randomization process, which provides a benchmark to compare the outputs of the encryption device. The uniform distribution is used as a reference for the entropy analysis, as it provides the maximum bound for the Shannon entropy. To test the behavior of random signals following these distributions, we consider the ensembles $\mathcal{S}^{(n);N}=\{\mathbf{x}^{(n);k,N}\}_{k=1}^{1000}$ and $\mathcal{S}^{(u);N}=\{\mathbf{x}^{(u);k,N)}\}_{k=1}^{1000}$ composed of 1000 normal and uniform randomly generated signals, respectively, for different dimensions, $N=\{5,10,\ldots,1000\}$. 

The truncated autocorrelation in~\eqref{chopped-xcorr} shall approximate a flat function for white noise signals, which, for $N$ finite dimension signals, is represented by a vector with an approximate null mean ($\mu\left(\widetilde{X}[\mathbf{x}]\right)$) and a standard deviation ($\sigma\left(\widetilde{X}[\mathbf{x}]\right)$) approaching zero as $N\rightarrow\infty$. To illustrate the latter, we compute the mean and standard deviation of the truncated autocorrelation of every element in the ensemble as $\mathcal{S}_{X;\mu}^{(n);N}:=\{\mu(\widetilde{X}[\mathbf{x}^{(n);k,N}])\}_{k=1}^{1000}$ and $\mathcal{S}_{X,\sigma}^{(n);N}:=\{\sigma(\widetilde{X}[\mathbf{x}^{(n);k,N}])\}_{k=1}^{1000}$, respectively. In this form, the behavior of the ensemble for each $N$ is revealed by computing the average of the mean of $\mathcal{S}_{X;\mu}^{(n);N}$ to get the main trend and the mean of $\mathcal{S}_{X;\sigma}^{(n);N}$ for the deviations around the main trend. The latter is shown in the blue-dashed area in Figure~\ref{fig:entropy1}\textbf{b}, where the expected tendency of the normal distribution is evident. For completeness, the same analysis was carried out for uniformly distributed signals (green-shaded).

In turn, the Shannon entropy $S(\mathbf{x})$ of a vector $\mathbf{x}$ provides a notion of uncertainty for probability distributions (see Methods section). Indeed, although a higher entropy does not necessarily imply higher randomness, one can still define a threshold for the entropy to identify a white-noise signal. For instance, the uniform distribution possesses the higher uncertainty, and thus maximum entropy, among all distributions with compact support~\cite{cover1999elements}. 
On the other hand, for Kronecker-delta-like distributions, the entropy is minimum (null) as certainty is absolute. Thus, the corresponding entropy shall be lower for a given white noise signal than that of uniform distributions. 

By using the previously introduced random ensembles of normal ($\mathcal{S}^{(n);N}$) and uniform ($\mathcal{S}^{(u);N}$) distributions, we can perform a statistical analysis based on the mean and standard deviation of the entropy for each ensemble across different signal sizes $N$. These results are depicted in Figure~\ref{fig:entropy1}\textbf{c}, where the solid and dashed curves denote the mean for normal and uniform ensembles, respectively. The shaded areas represent the corresponding standard deviations from the mean value for each distribution. The mean entropy is higher for uniform distributions, which is expected, as elements uniformly distributed are all equally likely and possess larger uncertainty. 

Thus, a given signal $\mathbf{x}$ is said to be white noise if the mean ($\mu\left( \widetilde{X}[\mathbf{x}] \right)$) and standard deviation ($\sigma\left( \widetilde{X}[\mathbf{x}] \right)$) of the truncated autocorrelation, and entropy $S(\mathbf{x})$ all lie in the normal distribution regions depicted in Figures~\ref{fig:entropy1}\textbf{b-c} for the corresponding size $N$. Note that if $\mathbf{x}=(0,\ldots,1,\ldots0)$, it follows that $\mu(\widetilde{X}[\mathbf{x}])=\sigma(\widetilde{X}[\mathbf{x}])=0$, which lies in the white-noise region but is clearly not a random signal. For these reasons and to rule out false positives, both conditions are enforced to conclude about the randomness of the signal in question. Remark that, for $N\lessapprox 15$, the entropy regions for both normal and uniform distributions are indistinguishable when analyzed using the entropy and autocorrelation criterion; thus, white noise is challenging to assess for relatively small-size signals.

Since the PIC produces optical complex-valued outputs, the white noise criteria shall be applied to the real and imaginary parts. In the current experimental setup, power measurements are gathered at the output, corresponding to the modulus square of the complex electric field, and we thus shall apply an equivalent criterion to the power. For simplicity and without loss of generality, we focus on the modulus $\vert \mathbf{x}\vert$. It is known that if the real and imaginary parts of $\mathbf{x}$ are normally distributed with mean zero and standard deviation one, the elements of $\vert \mathbf{x}\vert:=(\vert x_{1}\vert,\ldots,\vert x_{N}\vert)$ are distributed according to the Rayleigh distribution with scale parameter one~\cite{forbes2011statistical}. See upper-panel in Figure~\ref{fig:entropy1}\textbf{d}. From this, the autocorrelation of the $\vert \mathbf{x}\vert$ becomes linear with respect to the lag $\ell$ and anti-symmetric around $\ell=0$. Thus, the finite difference of the truncated autocorrelation, $\Delta_{\ell}\widetilde{X}[\vert\mathbf{x}\vert]:=(X_{2}-X_{1},\ldots,X_{N-2}-X_{N-1})$, also renders a flat distribution for $\ell>0$. See lower-panel in Figure~\ref{fig:entropy1}\textbf{d}. In this form, an equivalent criterion can be introduced for $\vert \mathbf{x}\vert$ based on $\Delta_{\ell}\widetilde{X}[\vert \mathbf{x}\vert]$ and the entropy $S(\vert\mathbf{x}\vert)$, which are respectively depicted in Figures~\ref{fig:entropy1}\textbf{e-f}. The analysis for the modulus of uniform distributions was excluded in the latter figure, as elements of such a distribution are already positive numbers. 



\subsection{Device randomness and encryption capabilities}
\begin{figure*}[htbp]
    \centering
    \includegraphics[width=0.85\textwidth]{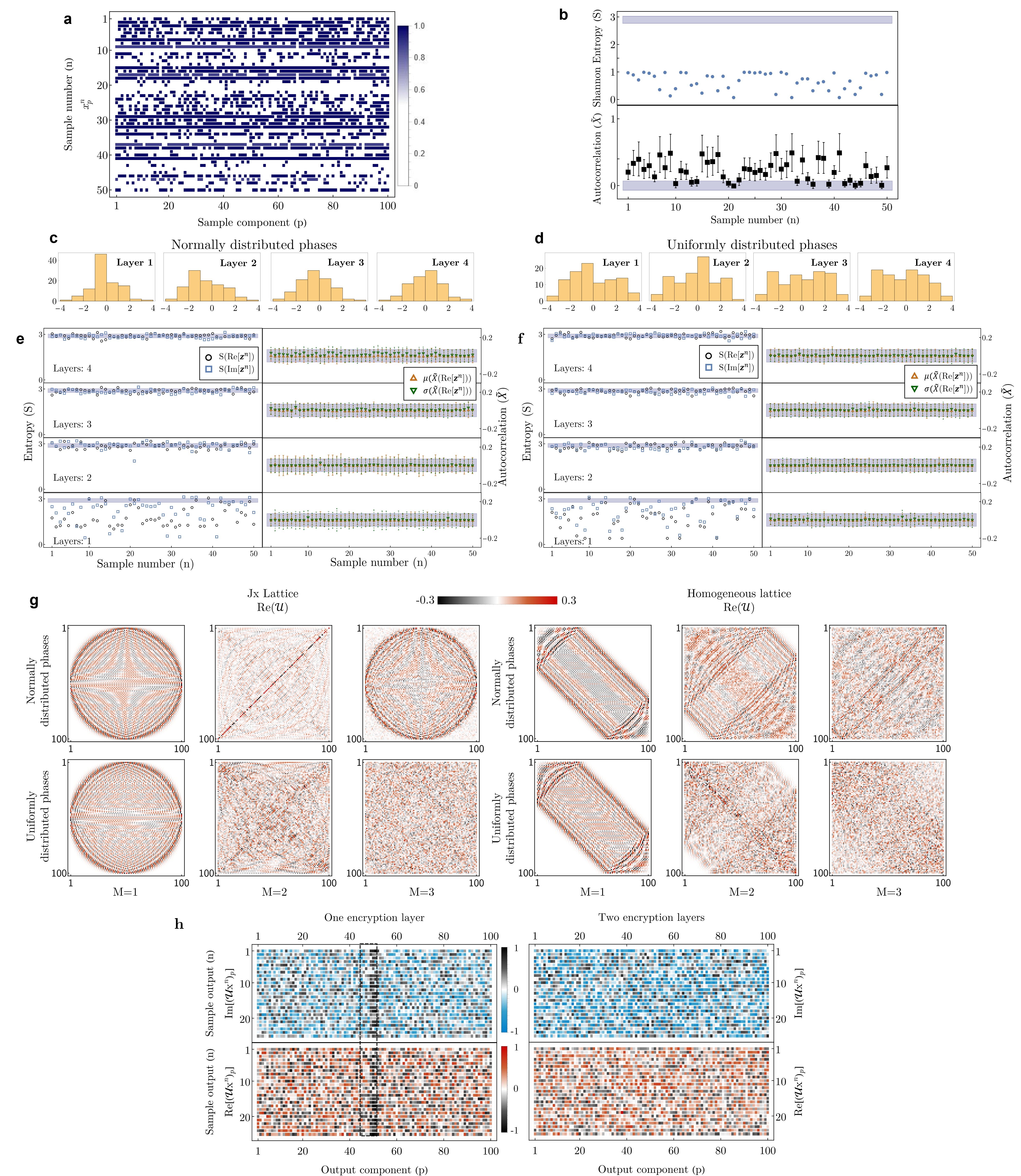}
    \caption{\textbf{Numerical simulations of the device randomness capabilities}. (a) Set of 50 randomized pulsed sample signals $\mathbf{x}^{n}$. (b) The corresponding values for the Shannon entropy (upper panel) and autocorrelation bars (lower panel). The shaded area denotes the region where white noise is expected. (c)-(d) Histograms of randomly generated phase shifters (encryption keys) taken from the normal (c) and uniform (d) distributions in the interval $(-\pi,\pi]$. (e)-(f) Entropy and truncated autocorrelation criteria for the samples randomized using normally (e) and uniformly (f) distributed keys. The random process uses $M=1, 2, 3, 4$ random-phase layers. (g) Real part of the transmission matrix $\mathcal{U}$ in Eq.~\eqref{encrypt} for $M=1,2,3$ layers and considering normally (upper panel) and uniformly (lower panel) distributed phases. For illustration, the Jx and homogeneous lattice were used as the mixing layers $F$. (h) Real and imaginary parts of the first 25 encrypted sample signals using normal keys combined with one (left) and two (right) encryption layers.}
    \label{fig:sparse}
\end{figure*}
The randomization capabilities of the proposed PIC are tested by first generating a set of input samples $\{ \widetilde{\mathbf{x}}^{n} \}_{n=1}^{50}$, where each sample $\mathbf{x}^{n}\in\mathbb{R}^{100}$, the components of which render a sparse signal generated from a sequence of random randomly placed unit pulses $\delta_{k,p}$, with $\delta_{k,p}$ the Kronecker delta function and $k,p\in \{ 1, \ldots,N=100 \}$. See Figure~\ref{fig:sparse}\textbf{a}. Despite the randomness in the generation of input sparse signals, they do not show any trace of white-noise behavior. This is done by testing the entropy and truncated autocorrelation criteria, as shown in Figure~\ref{fig:sparse}\textbf{b}, where the regions where white noise is expected (shaded areas) are highlighted. The entropy of every input sample lies below the expected values for white noise, whereas the mean and standard deviations deviate from the expected flat distribution for white noise in most cases. 

Therefore, a signal is deemed white noise if both entropy and truncated autocorrelation lie around the shaded corresponding regions. The requirement of both simultaneous criteria is better illustrated in sample No. 21, which has a low entropy but a perfect flat autocorrelation. This is because sample No. 21 is a single-unit pulse, the autocorrelation of which is easily proved to be flat, and thus the signal is not white noise, likewise, for other samples with a similar pattern. Thus, given that none of the samples fulfill the randomness criterion, we rule out the possibility that any trace of randomness eventually found at the PIC output is produced due to intrinsic randomness in the generation of input samples.

For the numerical tests performed in this section, the phase elements in each layer of the architecture are independently generated from either a normal or uniform distribution bounded to the interval $(-\pi,\pi)$. The corresponding histograms of the random phases used in the encryption process are illustrated in Figure~\ref{fig:sparse}\textbf{c-d} for up to four different layers. The latter allows analyzing the encryption capabilities of the output processed signals $\widetilde{\mathbf{z}}^{n}=\mathcal{U}\widetilde{\mathbf{x}}^{n}$ by inspecting the white noise behavior. To this end, the entropy and truncated autocorrelation are calculated for each sample output $\widetilde{\mathbf{z}}^{n}$ using both normal and uniform phases distributions as encryption keys, as illustrated in Figures~\ref{fig:sparse}\textbf{e-f}. Here, the numerical simulations are run considering $M=1,2,3,4$ encryption layers to showcase the effects of such added layers. Indeed, the entropy analysis shows that one encryption layer is insufficient to randomize all the input samples, even though the autocorrelation shows an almost flat distribution in each sample. We thus rule out the device with only one phase layer as a potential randomization device. 

In turn, when two phase layers are considered, the output of each sample signal produces higher entropy values, with only a few lying outside the region of white noise. Interestingly, the autocorrelation shows a higher standard deviation for normally distributed phases than those outputs encrypted with uniform keys. For more layers ($M>2$), both the entropy and autocorrelation criteria show no significant improvement as compared to $M=2$ layers. This numerical evidence allows for reducing the PIC size to $M=2$ layers without impacting random performance in any significant way. For completeness, the real part of the transfer matrices $\mathcal{U}$ is depicted in Figure~\ref{fig:sparse}\textbf{g} when sweeping the number of layers $M$ and using the Jx and homogeneous lattices as the passive mixing layers $F$. In analogy to the previous analysis, normally (upper panels) and uniformly (lower panels) distributed phases are also used here. It is clear that $M=1$ layers produce a transfer matrix resembling the original missing layer $F$ (see Supplementary Material S1), whereas $M=2$ layers wash out any such pattern. In both cases, the transfer matrices associated with uniformly distributed phases show a better random pattern than the normally distributed case. Thus, when operated with uniformly distributed phases, the PIC output produces a better random process.  

The randomization PIC can be further exploited to operate as an encryption device. This is done by treating the input signal as the vector to be encrypted, the phase distribution as a set of encryption keys, and the randomized output as the encrypted vector. In this procedure, the phase distribution has to be recorded and stored for subsequent decryption tasks. Indeed, the decryption process can be inverted, as the transformation operator $\mathcal{U}$ is unitary. Thus, the encrypted signal $\bf{x}_{out}$ can be cast back to its original form by injecting it into either the inverse operator $\mathcal{U}^{\dagger}$ or by reverting the device ports and using the conjugated phases. This inversion process can be done only if the original sets of phases $\phi_{n}^{(m)}$ used to randomize the input signal are known. For more details on the inversion process, see Supplementary Materials S2.

The real and imaginary parts of the first 25 normally encrypted samples are shown in Figure~\ref{fig:sparse}\textbf{g} for one and two encryption layers. Here, one can corroborate that one-layer encryption produces signals whose real and imaginary parts tend to pile up around the middle signal component 50 (dashed-rectangle in Figure~\ref{fig:sparse}\textbf{g}), as predicted from the entropy analysis. Indeed, such a tendency vanishes when the second encryption layer is added, producing signals spread across all the components. It is worth remarking that, for power measurements, the second phase layer in the architecture in Figure~\ref{fig:block} does not modify the readout at the power detectors. Thus, we can disregard the latter when performing power measurements. Nevertheless, the proposed PIC is flexible and compatible with phase measurement if the input and data collection stages of the present experimental run are changed by an optical vector analyzer. 


\subsection{Random passive layers}
Although the previous construction has shown a compact encryption device that can be scaled down to two phase layers without jeopardizing the encryption capabilities, it is always desirable to design a circuit with fewer elements. So far, the encryption is stored in the phase element, whereas the waveguide is fixed as a well-patterned unitary matrix. Thus, the device randomness can be enhanced by adding disorder into the waveguide arrays. The waveguides cannot be tuned once manufactured, but their pattern highly affects the output. It has been shown in~\cite{zelaya2024goldilocks} that random matrices generated from the Haar measure serve as passive layers in universal unitary interlaced architectures. The latter means that such matrices are dense enough to shuffle the elements of an input vector and render it into an arbitrary new one, provided that enough layers are available. For the encryption tasks, the encryption two-layered architecture should work when operated with random Haar matrices instead of a predefined lattice model, provided that the former matrices are dense.

The additional random element, namely the passive unitary layer $F$, is expected to increase the potential randomness at the encryption device output. To analyze the latter, we add the random symmetric deformation $R$ to the $J_x$ Hamiltonian $H$ so that the perturbed encryption device $\mathcal{U}(\delta)$ and the evolution through the perturbed waveguide array $F(\delta)$ read, respectively, as
\begin{equation}
\mathcal{U}(\delta)=P^{(2)}F(\delta)P^{(1)}F(\delta), \quad F(\delta)=e^{iz(H+\delta R)} ,
\label{U_enc_delta}
\end{equation}
where $\delta$ the perturbation strength parameter, and $R\in\mathbb{C}^{N\times N}$ a random matrix with elements taken from the normal distribution with mean $\mu=0$ and standard deviation $\sigma=1$; i.e., $\mathcal{N}(\mu=0,\sigma=1)$. Without loss of generality, the perturbation strength is considered as $\delta>0$. If the strength parameter $\delta\ll max(\kappa_{n})$, the deformation can be considered as a perturbation of the original $J_x$ lattice. For larger $\delta$, the overall effect of $R$ will overcome that of $J_x$, rendering a random matrix. This is handy as we can study the effects of small perturbation on the waveguide array, and also analyze the encryption capabilities of the device when the passive element is random in nature.

\begin{figure*}[t]
    \centering
    \includegraphics[width=0.8\textwidth]{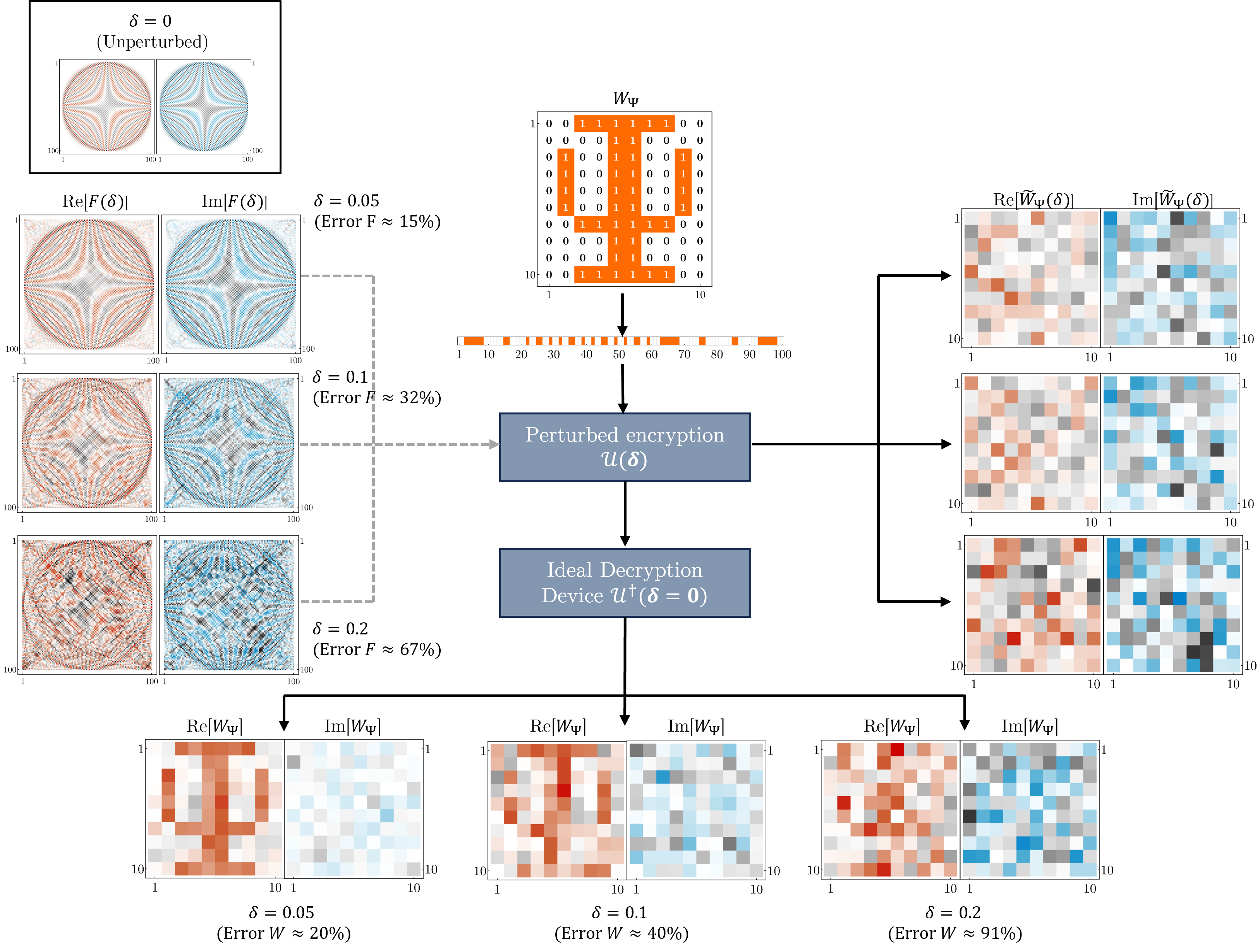}
    \caption{\textbf{Random defects and encryption capabilities.} (left panel) Real and imaginary parts of the perturbed DFrFT matrix $F(\delta)$ for $\delta=0.05,0.1,0.2$. (Top panel) Testing image $W$ used for encryption with perturbed DFrFT $F(\delta)$. (right panel) The corresponding encrypted images. (Bottom panel) Decrypted images using an ideal decryption device $\mathcal{U}^{\dagger}(\delta=0)$.}
    \label{fig:encrypted}
\end{figure*}

In order to measure the overall effect of the perturbation parameter $\delta$, it is more convenient to compute the percentage error introduced to $F(\delta)$ with respect to the ideal $J_x$ lattice; i.e., $E_{F}(\delta)=\left( \Vert F(\delta)-F \Vert/\Vert F \Vert\right)\times 100\%$. By considering 100 random perturbations for each $\delta$, one finds that, on average, the perturbations $\delta=0.05,0.1,0.2$ induce errors on $F$ around $E_{F}=15\%, 32\%, 67\%$, respectively. The overall effect of such perturbations on the real and imaginary parts of $F(\delta)$ are shown in Figure~\ref{fig:encrypted} (left panel). To test the effects of such perturbation on the encryption and decryption scheme, we consider a multi-stage setup. First, take a $10\times 10$ pixel image and reshape it into a 100-dimensional one-dimensional vector, which is encrypted using~\eqref{U_enc_delta} for $\delta=0.05,0.1,0.2$, with $N=100$. The encrypted vector is then decrypted using the same keys through the unperturbed decryption device $\mathcal{U}_{enc}^{\dagger}$. Indeed, the original source image is recovered when encrypted using $\delta=0$, whereas deviations are expected for $\delta\neq 0$. For $\delta\neq 0$, the decrypted images show an error due to the defects on the $F(\delta)$ layers. For instance, for $\delta=0.05$ ($E_F\approx 15\%$), the error in decrypting the image is approximately $E_{\widetilde{W}_{\Psi}}\approx 20\%$. Figure~\ref{fig:encrypted} (lower panel) shows the real and imaginary parts of the decrypted images, where it is clear that, for $\delta=0.05$, the real part still resembles the source image and the large error is due to the imaginary part components. In turn, for $\delta\leq 0.1$, the real part of the decrypted images is indistinguishable. 
%

\section{Discussion}
The randomization PIC has been tested under the proposed demultiplexing scheme, showing the expected performance. Among the potential issues present in the experimental run, the thermal heaters might induce undesired effects due to thermal cross-talk and thermal stabilization. The first effect is not relevant, as the phase shifters are randomly assigned from distributions, and thermal cross-talk will also contribute to the overall random pattern. Despite the latter, a set of pre-programmed phases will induce the same behavior in the PIC. This holds as long as thermal stabilization time is reached, which is the most critical factor to reach reproducibility. Thus, the experimental data was gathered after allowing a long enough time window $\tau$ between measurements after powering up all the metal heaters. The proposed PIC is flexible enough to be scaled up to a larger number of channels if so required, the design of which would only require that the transfer matrix of the passive mixing layer $F$ fulfills the required density criterion~\cite{zelaya2024goldilocks}. 

The proposed PIC successfully demonstrated the capability to generate the necessary random pattern and encryption features using a two-layer design. Despite adding a third layer not significantly improving the white noise of the encrypted signals, the extra key combinations used in the encryption process make the output more difficult to reverse-engineer without prior knowledge of the keys. In this regard, the proposed design can be adjusted whenever compactness or security is the final goal. There are reports in the literature for all-optical encryption devices in free-space configurations using random phase masks~\cite{refregier1995optical,unnikrishnan2000optical}. The latter includes a two-lens configuration combined with two statistically independent white noise phase planes, creating an encrypted image. The device introduced in the present work provides an equivalently compact and on-chip solution, which can be further exploited as an optical image encryption technique by following a similar demultiplexing procedure as the one discussed above. 

The measurements presented in this work mainly focus on power detection, but this is due to limitations in the experimental setup rather than the performance of the PIC itself. In fact, the proposed PIC is versatile and can support phase measurement if an optical vector analyzer is implemented or further interferometry is performed to extract the output phases. Furthermore, the proposed device can work as a versatile and controllable platform for investigating equivalent random and disordered wave systems~\cite{mafi2019disordered,mafi2021review}, where the disorder can be tuned on real-time to induce the desired effects. Recent studies have demonstrated that random photonic devices can serve as an effective tool for producing operations equivalent to Haar-random matrices. This is a crucial requirement for boson sampling in quantum computing tasks~\cite{russell2017direct,clementi2023programmable}. The latter is achieved by changing to other material platforms, such as silicon nitride (Si3N4), which are more suitable for single photon transport. In this regard, the intrinsic nonlinearity of the waveguide core can induce photon entanglement across a waveguide array~\cite{belsley2020generating}. 

\section{Methods}
\subsection{Material platform}
The randomization operation in the photonic integrated circuit (PIC) is carried out using a silicon-on-silica platform. A passive layer, denoted as $F(\alpha)$, is implemented using waveguide arrays that, based on the coupled-mode theory, facilitate the unitary wave evolution responsible for mixing a single excitation channel across all the waveguides. The waveguides are constructed with a silicon core (Si) surrounded by a silica cladding (SiO2) with refractive indices of $n_{Si}=3.47$ and $n_{SiO2}=1.4711$ at room temperature (293 K). In terms of geometry, we have a waveguide with a transverse rectangular shape featuring 500 nm width and 220 nm thickness. This configuration enables the waveguide to support a fundamental quasi-TE0 mode and a quasi-TM mode when operated at 1550 nm wavelengths. The PIC is specifically designed to operate on the quasi-TE0 mode, and 8-degree TE grating couplers are employed to effectively couple the right mode from the injection fiber into the PIC.

For the 5-channel PIC, the waveguide array is designed such that it exhibits symmetry around the middle waveguide. The spacing between the outer-most and middle waveguides is 233 nm and 210 nm, respectively, whereas the coupling length is set to 62 $\mu m$. The phase shifters are implemented using Ti/W alloy as heaters and are connected to the probing pads using bi-layer TiW/Al electrical traces. The probing pads are wire-bonded to a PCB for electrical access to the phase shifters. 


\subsection{Entropy and autocorrelation estimation}
While the current experimental setup is designed for capturing power measurements exclusively, the suggested PIC has the capability to acquire phase information by adjusting the data collection stage. This would allow for the collection of complex-valued signals, requiring the application of the white-noise criterion to both the real and imaginary parts, as well as the modulus of the signal, such as the case presented in the main text. 

Particularly, the autocorrelation $\ell$ component (lag) of the $\mathbf{X}[\mathbf{x}]$ is defined as $X_{\ell}[\bf{x}]:=(T_{\ell}\bf{x},\bf{x})$, with $T_{\ell}$ the translation operator, $(\bf{x},\boldsymbol{y})$ the Euclidean inner product in $\mathbb{R}^{N}$, and $\ell\in\{-N-1,\ldots,N-1\}$ the position of the lagged signal. In turn, to compute the Shannon entropy of either component of the signal, $\mathbf{x}\in\mathbb{R}^{N}$, it is first required to extract and normalize the related histograms $\{p_{j}\}_{j=1}^{M}$ using $M=\lfloor\sqrt{N}\rfloor$ bins. This allows producing the corresponding probability distribution to compute the Shannon entropy
\begin{equation*}
S=-\sum_{j=1}^{M}p_{j}\textnormal{log}_{2}p_{j} .
\end{equation*}


\bibliography{biblio}

\begin{backmatter}
\bmsection{Funding} This project was supported by the U.S. Air Force Office of Scientific Research (AFOSR) Young Investigator Program (YIP) Award\# FA9550-22-1-0189, the Defense University Research Instrumentation Program (DURIP) Award\# FA9550-23-1-0539, and by the National Science Foundation NSF-BSF Award\# DMR-2235381.

\bmsection{Data availability} The datasets generated during and/or analyzed during the current study are available from the corresponding author on reasonable request.

\bmsection{Supplemental document}
See the Supplementary Materials document for supporting content. 

\bmsection{Decelerations} Patent pending.

\end{backmatter}


\end{document}